\documentclass[
prd,
aps,
preprint,
nofootinbib,
floats]{revtex4}      

\usepackage{graphicx}
\usepackage{ifthen}
\usepackage{amsmath}
\usepackage{mathrsfs}
\usepackage{bbm}
\usepackage[normalem]{ulem}
\usepackage[usenames]{color}

\newcommand{\half}{\frac{1}{2}}

\newcommand{\fourth}{\frac{1}{4}}

\newcommand{\eighth}{\frac{1}{8}}

\newcommand{\abs}[1]{ \left| \, {#1} \right| }

\DeclareMathOperator{\real}{Re}
\DeclareMathOperator{\imag}{Im}

\newcommand{\inp}[2][0cm]{ \left( #2 \parbox[h][#1]{0cm}{} \right) }
\newcommand{\inb}[2][0cm]{ \left[ #2 \parbox[h][#1]{0cm}{} \right] }

\newcommand{\inap}[2][0cm]{ \left< {#2} \parbox[h][#1]{0cm}{} \right> }
\newcommand{\nop}[1]{\left.{#1}\right.}

\DeclareMathOperator{\Tr}{Tr}
\DeclareMathOperator{\col}{col}
\DeclareMathOperator{\row}{row}
\DeclareMathOperator{\diagOp}{diag}
\newcommand{\diag}[1]{\diagOp\inp{#1}}

\newcounter{itemnum}
\newenvironment{olist}
{%
\begin{list}%
	{(\roman{itemnum})}%
	{
	\usecounter{itemnum}}%
}%
{%
\end{list}
}%

\newcommand{\eqn}[1]{\eqref{Eq:#1}}
\newcommand{\tbl}[1]{TABLE~\ref{Table:#1}}
\newcommand{\NSUSY}[1][\normalsize]{\mbox{#1\sout{\raisebox{-1pt}{SUSY}}}}
\newcommand{\MP}{M_{P\ell}} 
\newcommand{\eq}[1]{Eq.~\eqref{Eq:#1}}
\newcommand{\spp}{\mbox{\tiny$++$}}	
\newcommand{\smm}{\mbox{\tiny$--$}}	
\newcommand{\DBarCpp}{\bar{\Delta}^{c\spp}}
\newcommand{\DBarCp}{\bar{\Delta}^{c +}}
\newcommand{\DBarCz}{\bar{\Delta}^{c \,0}}
\newcommand{\DCmm}{\Delta^{c\smm}}
\newcommand{\DCm}{\Delta^{c -}}
\newcommand{\DCz}{\Delta^{c \,0}}
\newcommand{\phiUp}{\phi_u^+}
\newcommand{\phiUz}{\phi_u^0}
\newcommand{\phiDz}{\phi_d^0}
\newcommand{\phiDm}{\phi_d^-}
\newcommand{\SCmix}{M_{\Phi \! \Delta^{c \! \pm}}^2}
\newcommand{\IThreeR}{$G_{211}$}
\newcommand{\E}[1]{\times 10^{#1}}

\begin{document}             
\preprint{\vbox{
\hbox{UMD-PP-05-053} }}
\title{\Large  Predicting the Seesaw Scale in a Minimal Bottom-Up Extension of MSSM}
\author{\bf R.N. Mohapatra, N. Setzer and S. Spinner }

\affiliation{ Department of Physics and Center for String and Particle
Theory, University of Maryland, College Park, MD 20742, USA}

\date{November, 2005}

\begin{abstract}
We analyze a minimal bottom-up seesaw scenario where we require the theory to satisfy three phenomenological conditions:
(i)~it is supersymmetric; 
(ii)~it has a local $B-L$ symmetry as part of the  $SU(2)_L \times SU(2)_R \times U(1)_{B-L}$ gauge theory to implement the seesaw mechanism and 
(iii)~$B-L$ symmetry breaking is such that it leaves R-parity unbroken giving a naturally stable dark matter. 
We show that in such a theory, one can predict the seesaw scale for neutrino masses to be $M_R\simeq 
\sqrt{M_{SUSY}\MP}\simeq 10^{11}$ GeV. We show that the ground state with this property is a stable minimum and is lower than possible electric charge violating minimum in this theory. Such models in their generic version are known to predict the existence of a light doubly charged Higgs boson and Higgsino which can be searched for in collider experiments. We give expressions for their masses in this minimal version. We then indicate how one can get different expectation values for the MSSM Higgs doublets in the theory required to have realistic quark masses.
\end{abstract} 

\maketitle

\section{Introduction}

One of the simplest ways to understand the small neutrino masses is to
use the seesaw mechanism where one adds a right-handed neutrino to the
standard model. This allows a Higgs coupling that leads to a
mass connecting the right- and the left-handed neutrinos, but this is much 
too large to be acceptable due to naturalness arguments that dictate this 
mass $m_D$ is of order of the known quarks and leptons.  However, the fact 
that the right-handed neutrino is a standard model
singlet allows one to write a large Majorana mass $M_R$ for it
leading to a small effective Majorana mass for the left-handed neutrino 
given by $m_\nu\sim -\frac{m^2_D}{M_R} \ll m_{e,u,d}$. This is known as 
the seesaw mechanism\cite{seesaw}. 

The seesaw mechanism raises the following questions:
\newcounter{item}
\begin{olist}
\item is there a natural way for the right-handed neutrino to appear in 
the theory rather than just being added to the standard model by hand?

\item how large is the seesaw scale $M_R$? In particular, why is $M_R\ll 
\MP$  as required by observations? Can we 
predict the value of $M_R$?
\end{olist}

The answers to these questions are connected. To answer the second 
question, one may start with the observation that 
the Majorana masses of the right-handed neutrinos break the $B-L$ 
symmetry, and if $B-L$ is a gauge symmetry of Nature\cite{marshak}, then 
that will explain why $M_R\ll \MP$. It is known that if the weak gauge 
group beyond the standard model is either $SU(2)_L\times 
U(1)_{I_{3R}}\times U(1)_{B-L}$ (\IThreeR) or the left-right symmetric 
group $SU(2)_L\times SU(2)_R \times U(1)_{B-L}$, then the right-handed 
neutrino appears naturally due to reasons of anomaly cancellation. So, 
$B-L$ naturally explains both the seesaw scale and 
the presence of $\nu_R$.

None of these considerations, however, tell us what the magnitude of the 
seesaw scale $M_R$ is. Observations tell us that if we want to understand 
the atmospheric neutrino data in a seesaw scheme we get $M_R\sim 10^{11}-10^{15}$ GeV, depending on the 
magnitude of the Dirac mass of the neutrino for the third generation 
(between 1-100 GeV). The higher 
value is tantalizingly close to the conventional grand unification scale 
in supersymmetric theories. As a result, in GUT theories such 
as $SO(10)$, one can identify $M_R$ with the scale of grand unification. 
Yet such theories allow many different values for $M_R$ while 
remaining consistent with the grand unification of 
couplings\cite{chang}, depending on how the symmetry is broken and the 
choice of Higgs multiplets, which prevents this connection between 
$M_R$ and grand unification from being unique. Nonetheless, simple  one 
or two step symmetry breaking $SO(10)$ models have provided a compelling 
class of models for studying the consequences of the seesaw mechanism for 
neutrino masses and mixings and need to be taken very seriously.

In this paper, we take an alternative point of view to the understanding of 
neutrino masses by making a minimal extension of the standard model to the 
supersymmetric left-right model (SUSYLR)\cite{susylr} or \IThreeR.  For the 
SUSYLR models we wish to consider, there are three defining phenomenological 
conditions:
\begin{olist}
\item it is supersymmetric;

\item it has a local $B-L$ symmetry as part of the gauge group 
$SU(2)_L\times SU(2)_R \times U(1)_{B-L}$ so that one can
implement the seesaw mechanism and

\item $B-L$ symmetry breaking is such that it leaves R-parity unbroken 
so that there is a naturally stable dark
matter.

\end{olist}

We will show in this paper that if in addition to the above, we add to 
this model an R-symmetry (not to be confused with R-parity, which is a 
property of the model regardless of the R-symmetry assumed), then it will 
predict $M_R\simeq \sqrt{M_{SUSY}\MP}\sim 10^{11}$ GeV which is of the 
right order of magnitude to be the seesaw scale. This is the main result 
of our paper and we believe that this is of interest since it determines 
the seesaw scale from first principles without the assumption of grand 
unification.

One must be careful with both the R-symmetry and the R-Parity because
R-symmetry restricts the number of parameters in the superpotential and
R-parity conservation allows a stable charge violating vacuum; therefore,
it is not obvious \emph{a priori} that this model has a stable, electric 
charge conserving vacuum. In fact, it has been shown that if we take only 
renormalizable terms in the theory, the R-parity must break in the 
electric charge conserving vacuum\cite{kuchi}.  We demonstrate that in our
model (which includes non-renormalizable terms) there exists, within the
parameter space, a charge conserving vaccum lower than the charge violating
one.

Working, then, in the context of the charge conserving vacuum, our model 
contains light particles in addition to those found in the standard model.  
Among these are two unstable neutral scalar bosons and a neutral fermion 
in the 100 GeV mass range.  Unfortunately these particles have no 
experimental implications because they will be difficult to produce 
in the laboratory due to their weak coupling to standard model particles.

Yet there are light particles in our model that do have an experimental
consequence: these are light doubly charged particles with masses in the
100 GeV to TeV range and they are present in all SUSYLR models\cite{melfo}
(even those without the R-symmetry we impose). Phenomenological implications 
of these particle has been studied in \cite{dutta} and their experimental 
search has been conducted \cite{doubly} at LEP, CDF and D0 experiments 
and can be searched at the LHC. Low energy consequences of the doubly 
charged Higgs bosons have also been studied extensively\cite{dc}. The present 
lower limits on their masses is around 115 GeV or so depending on which channel 
the particle has been searched for.

The details of the above discussion will be presented in the remaining of the
paper in following order:
in sec.~\ref{Sec:Gen Model}, we present our minimal 
$SU(2)_L\times SU(2)_R \times U(1)_{B-L}$ model where the seesaw scale 
is predicted as the geometric mean of the weak scale and the Planck scale 
and analyze the ground state of the theory.  We then proceed to find the 
effective low-energy theory, verify that it contains the minimal 
supersymmetric model, and address the breaking of $SU(2)_L \times 
U(1)_Y$.  The end of section \ref{Sec:Gen Model} verifies that the 
ground state of the theory conserves electric charge.  In sec.~\ref{Sec:Mass} 
we present the masses of the Higgs bosons and check that there are no 
tachyonic states; additionally, we show that there are TeV scale doubly 
charged fermions and bosons in this theory---confirming earlier results 
derived for nonminimal models \cite{melfo}.  
In sec.~\ref{Sec:Grand Unification}, we discuss possible grand 
unification prospects for the model.

\section{Theoretical Model And the Seesaw Scale}
\label{Sec:Gen Model}

The quarks and leptons in this model are assigned to the representations 
of the group as shown in Table \ref{Table:QM Numbers}.  We assume that 
 parity symmetry is broken at a high scale so that the left handed 
 partner of the $\Delta ^c$ and $\bar{\Delta}^c$ are not included in the 
theory.

\begin{center}
\begin{table}[!ht]
\begin{tabular}{|c|c|}
\hline\hline
Fields			& $SU(2)_L \times SU(2)_R  \times U(1)_{B-L}$	\\
\hline
$Q$			& $(2, 1, +\frac{1}{3})$			\\
$Q^c$			& $(1, 2, -\frac{1}{3})$			\\
$L$			& $(2, 1, -1)$					\\
$L^c$			& $(1, 2, +1)$					\\
$\Phi$			& $(2, 2, 0)$					\\
$\Delta^c$		& $(1, 3, -2)$					\\
$\bar{\Delta}^c$	& $(1, 3, +2)$					\\
\hline\hline
\end{tabular}
 \caption{Assignment of the fermion and Higgs fields to the 
representation of the left-right symmetry group.}
\label{Table:QM Numbers}
\end{table}
\end{center}

For the sake of clarification, we note that the various fields transform 
under $SU(2)_L\times SU(2)_R$ as follows:
\begin{align*}
Q & \rightarrow U_L Q						& 
Q^c & \rightarrow U_R Q^c					&  
L & \rightarrow U_L L						&
L^c & \rightarrow U_R L^c					\\ 
\Delta^c & \rightarrow U_R \Delta^c U_R^\dagger			&
\bar{\Delta}^c & \rightarrow U_R \bar{\Delta}^c U_R^\dagger	&
\Phi_a  & \rightarrow U_L \Phi_a U_R^\dagger
\end{align*}
Given those transformations, the fields may be written as
\begin{align} \notag
Q & = \begin{pmatrix}u \\ d \end{pmatrix} 
&
Q^c & = \begin{pmatrix}d^c \\ -u^c \end{pmatrix}
&
L & = \begin{pmatrix}\nu \\ e \end{pmatrix}
&
L^c & = \begin{pmatrix}e^c \\ -\nu^c \end{pmatrix}
\\ \notag
\Delta^c 
	& = \begin{pmatrix}
		\frac{\DCm}{\sqrt2}		& \DCz			\\
		\DCmm				& - \frac{\DCm}{\sqrt 2}
	    \end{pmatrix}
&
\bar{\Delta}^c 
	& = \begin{pmatrix}
		\frac{\DBarCp}{\sqrt2}	& \DBarCpp			\\
		\DBarCz			& - \frac{\DBarCp}{\sqrt 2}
	    \end{pmatrix}
&
\Phi & = \begin{pmatrix}
	\phiDz	& \phiUp	\\
	\phiDm	& \phiUz
	\end{pmatrix}
\end{align}
where the neutral fields can be split into their real and imaginary parts by
\begin{align} \notag
\label{Eq:Neutral Real/Imag}
\DCz	& = \frac{1}{\sqrt{2}} \inp{\real{\DCz} + i \imag{\DCz}}
&
\DBarCz	& = \frac{1}{\sqrt{2}} \inp{\real{\DBarCz} + i \imag{\DBarCz}}
\\
\phiUz	& = \frac{1}{\sqrt{2}} \inp{\real{\phiUz} + i \imag{\phiUz}}
&
\phiDz	& = \frac{1}{\sqrt{2}} \inp{\real{\phiDz} + i \imag{\phiDz}}
\end{align}

The superpotential is given by
\begin{equation}
\label{Eq:SuperW}
W = W_Y + W_H
\end{equation}
where
\begin{equation}
\label{Eq:SuperW Terms}
\begin{aligned}[b]
W_Y	& =	  i h Q^T \tau_2 \Phi Q^c
	  	+ i h^\prime L^T \tau_2 \Phi L^c
		+ i f_c L^{cT} \tau_2 \Delta^c L^c
	\\
W_H	& =	\frac{\lambda_A}{\MP} \inp{ \Tr\inp{\Delta^c \bar{\Delta}^c} }^2
	 	+ \frac{\lambda_B}{\MP} \Tr\inp{\Delta^c \Delta^c} \Tr\inp{\bar{\Delta}^c \bar{\Delta}^c}
	\\
	&  \quad {} + \frac{\lambda_\alpha}{\MP} \Tr\inp{\Delta^c \bar{\Delta}^c} \Tr\inp{\Phi^T \tau_2 \Phi \tau_2}
\end{aligned}
\end{equation}

Several comments regarding this superpotential are in order:

\begin{olist}
\item There are no bilinear terms in the fields $\Delta^c$ or $\Phi$.  
This property is the one that permits the prediction of the seesaw  scale, 
and the lack of such terms can always be accomplished by requiring an 
appropriate R-symmetry in the theory; e.g., 
${(\bar{\Delta}^c, \Delta^c,  \Phi) \rightarrow e^{i\pi/2}(\bar{\Delta}^c,  \Delta^c, \Phi)}$; 
$(L,L^c,Q,Q^c)\rightarrow e^{-i\pi/4}(L,L^c,Q,Q^c)$.
\item There is no distinct  $\Tr\inp{\Delta^c \tau_2 \Phi^T \tau_2 \Phi 
\bar{\Delta}^c}$ term like the one that occurs in the non-supersymmetric 
theory.  This is due to the  fact that a superpotential must be 
holomorphic in the fields---excluding the appearance of a 
$\Phi^\dagger$.  Thus, the only invariant combination involving both the 
right-handed $\Delta$'s and the $\Phi$'s must be of the form $\Phi^T 
\tau_2 \Phi$ to satisfy $SU(2)_L$ invariance; the transpose then forces 
another $\tau_2$ to be involved.  This, coupled with the fact that 
$\Phi^T \tau_2 \Phi \tau_2 = \mathbbm{1} \cdot \det \Phi$, gives that 
$\Tr\inp{\Delta^c \tau_2 \Phi^T \tau_2 \Phi \bar{\Delta}^c} = \half 
\Tr\inp{\Delta^c \bar{\Delta}^c} \Tr\inp{\Phi^T \tau_2 \Phi \tau_2}$
\end{olist}

From Eqs.~\eqn{SuperW} and \eqn{SuperW Terms} we can write down the 
Higgs potential taking the supersymmetry breaking terms into account as 
follows:
\begin{equation}
\label{Eq:potential}
V(\Phi, \Delta^c, \bar{\Delta}^c)
	= V_F + V_D + V_{\NSUSY[\scriptsize]}
\end{equation}
where
\begin{align} \notag
\label{Eq:V F Terms}
V_F	& =	  \frac{4 \lambda_A^2}{\MP^2}
			\abs{ \Tr\inp{\Delta^c \bar{\Delta}^c} }^2
			\inp{  \Tr \abs{\Delta^c}^2 + \Tr \abs{\bar{\Delta}^c}^2 }
	\\ \notag
	& \quad {}
		+ \frac{4 \lambda_B^2}{\MP^2}
			\inb{
			 	  \abs{ \Tr\inp{\bar{\Delta}^c \bar{\Delta}^c} }^2 \Tr\abs{\Delta^c}^2
			 	+ \abs{ \Tr\inp{\Delta^c \Delta^c} }^2 \Tr\abs{\bar{\Delta}^c}^2
			}
	\\ \notag
	& \quad {}
		+ \frac{\lambda_\alpha^2}{\MP^2}
			\abs{ \Tr\inp{\Phi^T \tau_2 \Phi \tau_2} }^2
			 \inp{ \Tr \abs{\Delta^c}^2 + \Tr \abs{\bar{\Delta}^c}^2 }
		+ \frac{4 \lambda_\alpha^2}{\MP^2}
			\abs{ \Tr\inp{\Delta^c \bar{\Delta}^c} }^2
			\Tr \abs{\Phi}^2
	\\ \notag
	& \quad {}
		+ \left[
		  \frac{4 \lambda_A \lambda_B}{\MP^2} \Tr\inp{\Delta^c \bar{\Delta}^c}
			\Bigl(
				  \Tr\inp{\Delta^c \Delta^c}^* 
			 		\Tr\inp{\bar{\Delta}^{c \, \dagger} \Delta^c}
			 	+ \Tr\inp{\bar{\Delta}^c \bar{\Delta}^c}^* 
			 		\Tr\inp{\bar{\Delta}^c \Delta^{c \, \dagger}}
			\Bigr)
		\right.
	\\ \notag
	& \quad {}
		+ \frac{2 \lambda_A \lambda_\alpha}{\MP^2}
			\Tr\inp{\Delta^c \bar{\Delta}^c}
			\Tr\inp{\Phi^T \tau_2 \Phi \tau_2}^*
			 \inp{ \Tr \abs{\Delta^c}^2 + \Tr \abs{\bar{\Delta}^c}^2 }
	\\
	& \left. \quad {}
		+ \frac{2 \lambda_B \lambda_\alpha}{\MP^2}
			\Tr\inp{\Phi^T \tau_2 \Phi \tau_2}^*
			\Bigl(
			 	  \Tr\inp{\Delta^c \Delta^c} \Tr\inp{\bar{\Delta}^c \Delta^{c \, \dagger}}
			 	+ \Tr\inp{\bar{\Delta}^c \bar{\Delta}^c} \Tr\inp{\bar{\Delta}^{c \, \dagger} \Delta^c}
			\Bigr)
		+ \mbox{c.c} \vphantom{\frac{2 \lambda_B \lambda_\alpha}{\MP^2}} 
		\right]
\end{align}
\begin{align}
\notag
\label{Eq:V D Terms}
V_D	& =	  \frac{g_R^2}{8} \sum_{j} 
			\inb{\Tr\inp{
				  2 \Delta^{c \, \dagger} \tau_j \Delta^c
			 	+ 2 \bar{\Delta}^{c \, \dagger} \tau_j \bar{\Delta}^c
				+ \Phi \tau_j \Phi^\dagger
			}}^2
		+ \frac{g_L^2}{8} \sum_j
			\inb{\Tr\inp{
				\Phi^\dagger \tau_j \Phi
			}}^2
	\\
	& \quad {}
		+ \frac{g_{BL}^2}{2} 
			\inb{ \Tr\inp{
				  \Delta^{c \, \dagger} \Delta^c
				- \bar{\Delta}^{c \, \dagger} \bar{\Delta}^c
			}}^2
\end{align}
\begin{align}
\notag
\label{Eq:V non SUSY}
V_{\NSUSY[\scriptsize]}
	& =	- m_{\Delta^c}^2 \Tr\inp{\Delta^{c \, \dagger} \Delta^c }
		- m_{\bar{\Delta}^c}^2 \Tr\inp{\bar{\Delta}^{c \, \dagger} \bar{\Delta}^c }
		- m_{\Phi}^2 \Tr\inp{\Phi^\dagger \Phi}
	\\ \notag
	& \quad	{}
		- \frac{Z_A m_{3/2}}{\MP} \inb{ \inp{\Tr\inp{\Delta^c \bar{\Delta}^c}}^2 + \mbox{c.c.}}
		+ \frac{Z_B m_{3/2}}{\MP} \inb{ \Tr\inp{\Delta^c \Delta^c}\Tr\inp{\bar{\Delta}^c\bar
		{\Delta}^c}+\mbox{c.c.}}
	\\
	& \quad {}
		- \frac{Z_\alpha m_{3/2}}{\MP}
 			\inb{ \Tr\inp{\Delta^c \bar{\Delta}^c} \Tr\inp{\Phi^T \tau_2 \Phi \tau_2} + \mbox{c.c.}}
\end{align}
with $V_F$ being the $F$-term contribution, $V_D$ the $D$-term, and 
$V_{\NSUSY[\scriptsize]}$ the SUSY breaking terms.

The minima equations, correct up to electroweak order (i.e.~neglecting 
terms of order $v_{wk}^3/\MP$ and smaller), are:
\begin{align}
\label{Eq:Delta Min Deltac} 
	  \half \inp{g_{BL}^2 + g_R^2} \inp{v_R^2 - \bar{v}_R^2}
	- \fourth g_R^2 \inp{\kappa_u^2 - \kappa_d^2}
	- m_{\Delta^c}^2 
	- \frac{Z_A m_{3/2}}{\MP} \bar{v}_R^2 
 	+ \frac{ \lambda_A^2 \bar{v}_R^2 }{ \MP^2 } \inp{2v_R^2 + \bar{v}_R^2}
	& = 0
\\
\label{Eq:Delta Min Deltabarc} 
	- \half \inp{g_{BL}^2 + g_R^2} \inp{v_R^2 - \bar{v}_R^2}
	+ \fourth g_R^2 \inp{\kappa_u^2 - \kappa_d^2}
	- m_{\bar{\Delta}^c}^2 
	- \frac{Z_A m_{3/2}}{\MP} v_R^2 
	+ \frac{ \lambda_A^2 v_R^2 }{ \MP^2 } \inp{v_R^2 + 2 \bar{v}_R^2}
	& = 0
\end{align}
\begin{multline}
\label{Eq:Phi Min Phiu} 
\left[
	- \fourth g_R^2 \inp{ v_R^2 - \bar{v}_R^2 }
	+ \eighth \inp{g_R^2 + g_L^2} \inp{ \kappa_u^2 - \kappa_d^2 }
	- m_\Phi^2
	+ \frac{ \lambda_\alpha^2 v_R^2 \bar{v}_R^2 }{ \MP^2 }
  \right] \kappa_u
\\
{} - v_R \bar{v}_R \left[
	  \frac{Z_\alpha m_{3/2}}{\MP} 
 	- \frac{ \lambda_A \lambda_\alpha \inp{v_R^2 + \bar{v}_R^2} }{\MP^2}
  \right] \kappa_d
  	 = 0
\end{multline}
\vspace{-6ex}
\begin{multline}
\label{Eq:Phi Min Phid} 
\left[
	\fourth g_R^2 \inp{ v_R^2 - \bar{v}_R^2 }
	- \eighth \inp{g_R^2 + g_L^2} \inp{ \kappa_u^2 - \kappa_d^2 }
	- m_\Phi^2
	+ \frac{ \lambda_\alpha^2 v_R^2 \bar{v}_R^2 }{ \MP^2 }
  \right] \kappa_d
\\
{} - v_R \bar{v}_R \left[
	  \frac{Z_\alpha m_{3/2}}{\MP}
 	- \frac{ \lambda_A \lambda_\alpha \inp{v_R^2 + \bar{v}_R^2} }{\MP^2}
  \right] \kappa_u
  	 = 0
\end{multline}
where we have taken the VEV's to be the real part of the neutral field; 
that is
\begin{align}
 v_R & \equiv \inap{\real{\DCz}}		& \bar{v}_R & \equiv \inap{\real{\DBarCz}}	&
 \kappa_u & \equiv \inap{\real{\phiUz}}		& \kappa_d & \equiv \inap{\real{\phiDz}}
\end{align}

 Considering only \eq{Delta Min Deltac} and \eq{Delta Min Deltabarc} for 
the moment, take 
\begin{align}
\label{Eq:Ang Param}
v_R &= v \sin \theta_R		&
\bar{v}_R &= v \cos \theta_R	&
\kappa_u &= \kappa \sin \beta	&
\kappa_d &= \kappa \cos \beta
\end{align}

 Now, the difference of the squares of $v_R$ and $\bar{v}_R$ must be of 
 order $v_{wk}^2$ (subtracting \eq{Delta Min Deltac} from \eq{Delta Min 
 Deltabarc} will reveal this), so $\theta_R$ must be near $\pi/4$.  
Therefore, let
\begin{equation}
\theta_R = \frac{\pi}{4} + \frac{\epsilon}{2}
\end{equation}
and expand to first order in $\epsilon$ (as we shall see, $\epsilon \sim v_{wk}/\MP$---so $\epsilon$ is quite small).  The sum of equations \eqn{Delta Min Deltac} and \eqn{Delta Min Deltabarc} yields a quadratic for $v^2$:
\begin{equation}
\label{Eq:vSqr Quadratic}
- \inp{ m_{\Delta^c}^2 + m_{\bar{\Delta}^c}^2 }
- \frac{ Z_A m_{3/2} }{ \MP } v^2
+ \frac{3}{2} \frac{\lambda_A^2 v^4}{\MP^2}
= 0
\end{equation}
and the difference gives an expression for $\epsilon$:
\begin{equation}
\label{Eq:epsilon}
\epsilon = 
\frac{ m_{\Delta^c}^2 - m_{\bar{\Delta}^c}^2 - \half g_R^2 \kappa^2 \cos 2\beta }{ \inp{g_{BL}^2 + g_R^2} v^2 }
\end{equation}

The solution to \eq{vSqr Quadratic}, 
\begin{equation}
\label{Eq:vSqr}
v^2 = 
\inp{
\frac	{  Z_A m_{3/2} + \sqrt{ \inp{Z_A m_{3/2}}^2 + 6 \lambda_A^2\inp{m_{\Delta^c}^2 + m_{\bar{\Delta}^c}^2} }  }
	{ 3 \lambda_A^2 }
}\MP
\end{equation}
 gives our prediction of the right breaking scale.  Since $Z_A \sim 
 \lambda_A \sim 1$ and $m_{\Delta^c} \sim m_{\bar{\Delta}^c} \sim m_{3/2} 
 \sim v_{wk}$, we get the result $v \simeq \sqrt{v_{wk} \MP}$.  This is 
the major result of the paper, which shows that we can determine the seesaw 
scale in terms of two other commonly assumed and well motivated scales in 
the theory; i.e., the Planck scale in four dimensions and the supersymmetry 
breaking scale (which is of the order of the weak scale to solve the gauge 
hierarchy problem). This gives for the seesaw scale $M_R\simeq v\sim 
10^{11}$ GeV. Using 
this and \eq{epsilon} then shows that $\epsilon \lesssim v_{wk}/\MP$.

 Now turn to \eq{Phi Min Phiu} and \eq{Phi Min Phid}---again using 
 \eq{Ang Param} and expanding to first order in $\epsilon$, their sum 
yields an expression for $\sin 2\beta$:
\begin{equation}
\label{Eq:sin 2 beta}
\sin 2\beta =
 \frac	{ \inb{\frac{Z_\alpha m_{3/2}}{\MP} - \frac{\lambda_A \lambda_\alpha v^2}{\MP^2}} v^2 }
	{ \frac{\lambda_\alpha^2 v^4}{2 \MP^2} - 2 m_\Phi^2 }
\end{equation}
and their difference---after using \eq{epsilon} and \eq{sin 2 beta}---gives
\begin{equation}
\label{Eq:cos 2 beta}
\cos 2\beta =
 \frac	{ \half \frac{g_R^2}{g_{BL}^2 + g_R^2} \inp{m_{\bar{\Delta}^c}^2 - m_{\Delta^c}^2} }
	{
		  \fourth \inp{ g_L^2 + \frac{g_{BL}^2 g_R^2}{g_{BL}^2 + g_R^2} }\kappa^2 
		+ \frac{ \lambda_\alpha^2 v^4}{ 2 \MP^2 } 
		- 2 m_\Phi^2 
	}
\end{equation}
%

Both \eq{sin 2 beta} and \eq{cos 2 beta} are consistant with any value of $\beta$ and constrain the parameter space once a value of $\beta$ has been specified.  It is also easy to make an analogy between them and the usual Minimal Supersymmetric Standard Model (MSSM) results as we now do.

\subsection{Correspondence to MSSM}

 We  begin our discussion of the effective low energy theory with the 
 relationship between our parameters and those in the MSSM.  When 
 $SU(2)_R \times U(1)_{B-L}$ is broken the $\Tr\inp{\Delta^c 
 \bar{\Delta}^c} \Tr\inp{\Phi^T \tau_2 \Phi \tau_2}$ of \eq{SuperW Terms} 
 will yield a mass term for the $SU(2)_L$ doublets $\phi_u$ and $\phi_d$; 
 since these are basically the $H_u$ and $H_d$ of the MSSM, this must be 
the usual $\mu$ term.  We then have that
\begin{equation}
\abs{\mu} = \abs{ \frac{\lambda_\alpha v^2}{2 \MP} } \sim v_{wk}
\end{equation}
which is of the desired order of magnitude without any extra assumptions.

Similar reasoning yields that the SUSY breaking bilinear  term, $B$, 
will 
have a contribution resulting from the $Z_\alpha$ term in \eq{V non 
SUSY};  however, it will also receive a contribution from the $F$-term 
of \eq{V F Terms}---specifically the coefficient of $\lambda_A 
\lambda_\alpha$.  Together these give
\begin{equation}
B = \inb{ \frac{Z_\alpha m_{3/2}}{2 \MP} - \frac{\lambda_A \lambda_\alpha v^2}{2 \MP^2} }v^2
\end{equation}

Using the expressions for $B$ and $\mu$ and examining the minimization 
conditions given by \eq{Phi Min Phiu} and \eq{Phi Min Phid}, we can read 
off $m_{H_u}^2$ and $m_{H_d}^2$:
\begin{align}
m_{H_u}^2 & =	- m_\Phi^2 
		- \fourth g_R^2 \inp{v_R^2 - \bar{v}_R^2} 
		+ \eighth \frac{g_R^4}{g_{BL}^2 + g_R^2} \inp{\kappa_u^2 - \kappa_d^2}
\\
m_{H_d}^2 & =	- m_\Phi^2
		+ \fourth g_R^2 \inp{v_R^2 - \bar{v}_R^2}
 		- \eighth \frac{g_R^4}{g_{BL}^2 + g_R^2} \inp{\kappa_u^2 - \kappa_d^2}
\end{align}

 Here it is noticed that $m_{H_u}^2 \ne m_{H_d}^2$ despite the apparent 
 symmetry of the superpotential and the soft-breaking mass term.  This 
 splitting is due to the $D$-terms, which is reflected in the fact that 
 their difference is proportional to $g_R^2$.  Specifically, it is the 
 $D$-term involving $\tau_3$ (the ones involving $\tau_1$ and $\tau_2$ 
won't contribute because when the VEVs are placed in for $\Delta^c$ and 
$\bar{\Delta}^c$ these are zero) that gives a positive $\bar{v}_R^2 - 
v_R^2$ contribution to $m_{H_u}^2$ and a negative one to $m_{H_d}^2$.

Using \eq{epsilon} these expressions may be recast into the form
\begin{align}
m_{H_u}^2 & =	- m_\Phi^2 
		- \fourth \frac{g_R^2}{g_{BL}^2 + g_R^2} \inp{m_{\Delta^c}^2 - m_{\bar{\Delta}^c}^2}
\\
m_{H_d}^2 & =	- m_\Phi^2
		+ \fourth \frac{g_R^2}{g_{BL}^2 + g_R^2} \inp{m_{\Delta^c}^2 - m_{\bar{\Delta}^c}^2}
\end{align}

This form is advantageous because we now have that
\begin{equation}
m_{H_u}^2 - m_{H_d}^2  = \half \frac{g_R^2}{g_{BL}^2 + g_R^2} \inp{m_{\bar{\Delta}^c}^2 - m_{\Delta^c}^2}
\end{equation}
which, with the masses of the left-handed (standard model) particles
\begin{align}
\label{Eq:m Z}
m_Z^2 	& = \fourth \inb{ g_L^2 + \frac{g_{BL}^2 g_R^2}{g_{BL}^2 + g_R^2} } \kappa^2
\\
\label{Eq:m W} 
m_W^2	& = \fourth g_L^2 \kappa^2
\end{align}
allows us to write \eq{sin 2 beta} and \eq{cos 2 beta} in the enticing form
\begin{align}
\sin 2 \beta	& = \frac{2 B}{ 2\abs{\mu}^2 + m_{H_u}^2 + m_{H_d}^2 }
\\
\cos 2 \beta	& = \frac{ m_{H_u}^2 - m_{H_d}^2 }{m_Z^2 + 2\abs{\mu}^2 + m_{H_u}^2 + m_{H_d}^2 }
\end{align}
which are the usual expressions of the MSSM for breaking $SU(2)_L \times 
U(1)_Y$ down to $U(1)_{em}$.  

The interesting aspect of this result is that, provided $m_{\Delta^c}^2 
\ne m_{\bar{\Delta}^c}^2$, this model provides a means for $\tan \beta 
\equiv \frac{\kappa_u}{\kappa_d} \gg 1$.  This is an important feature 
because, as has already been noted for theories involving a single 
bidoublet\cite{bdm}, it allows the quarks and leptons to get realistic 
masses and mixings.  Since this model does not require additional 
particles to achieve  $\tan \beta \gg 1$ (as opposed to those previously 
discussed\cite{bdm1,shafi}), it is truly a minimal scheme.

\subsection{Charge Violation Consideration}

The above model is based on vacuum expectation values (vevs) that are 
consistant with the charge conserving vacuum.  However, it has been 
noted noted in earlier works that in \mbox{SUSYLR} models, the $\Delta^c$ 
fields may have a vev that breaks electric charge 
conservation\cite{kuchi} unless one breaks R-parity.  In this 
model though, the existence of non-renormalizable terms allow for the 
charge conserving vacuum to have a much lower ground state energy 
than the charge conserving one for large regions of the parameter 
space.  This ensures that the theory will spontaneously break into the 
phenomenologically viable vacuum, the charge conserving one.

To see this we can compare the ground state values of the two 
potentials, the charge  violating one (CV) and the charge conserving (CC) 
one.  The vevs for the  CC case have already been discussed, their 
analogues in the CV case are:
\begin{align} \notag
\inap{\Delta^c}
	& = \begin{pmatrix}
		0					& \frac{v_R}{\sqrt2}			\\
		\frac{v_R}{\sqrt2}			& 0
	    \end{pmatrix}
&
\inap{\bar{\Delta}^c}
	& = \begin{pmatrix}
		0					& \frac{\bar{v}_R}{\sqrt2}		\\
		\frac{\bar{v}_R}{\sqrt2}		& 0	    
	    \end{pmatrix}
\end{align}

The resulting ground state expressions, to order $v_R$, are:
\begin{align}
\inap{V}_{CV}
	& = 	- \half v^2 \inp{ m_{\Delta^c}^2 +  m_{\bar{\Delta}^c}^2 }
		+ \frac{\inp{Z_B - Z_A} m_{3/2} v^4}{2 \MP}
		+ \frac{\inp{\lambda_A + \lambda_B}^2 v^6}{\MP^2}
\\
\inap{V}_{CC}
	& = 	- v^2 \inp{ m_{\Delta^c}^2 + m_{\bar{\Delta}^c}^2 }
		- \frac{2 Z_A m_{3/2} v^4}{\MP}
		+ \frac{8 \lambda_A^2 v^6}{\MP^2}	
\end{align}

Where $v^2$ for the CC case was given in \eq{vSqr} and $v^2$ for the CV case is:

\begin{equation}
v^2 = 
\inp{
\frac{    \inp{Z_A -Z_B} m_{3/2} 
	+ \sqrt{ 
		\inp{
			  \inp{Z_A-Z_B} m_{3/2}}^2 
			+ 6 \inp{\lambda_A + \lambda_B}^2\inp{m_{\Delta^c}^2 
			+ m_{\bar{\Delta}^c}^2
		    } 
		}  
     }
     { 
     	6 \inp{\lambda_A + \lambda_B}^2 
     }
}\MP
\end{equation}

The crucial point here is that the CV ground state expression has a 
dependence on both $Z_B$ and $\lambda_B$, which do not appear in the CC 
expression.  This means that for sufficiently large values of these 
parameters, the CC ground state will be lower.  In the numerical 
analysis conducted in a later section, this will be taken 
into account and the difference between the two ground state values will 
be compared.

\section{Mass Spectrum and Numerical Analysis}
\label{Sec:Mass}

\subsection{Mass Spectrum}

Once the value of the minimization conditions and the values of the vevs 
have been determined, the mass spectrum can be explored to ensure that 
all the resulting physical Higgs bosons have positive mass squares. This is 
nontrivial because if too few terms are included in the superpotential, 
there is no \emph{a priori} guarantee that there 
is a stable minimum instead of a flat direction or an unstable minimum. 



We begin\footnote{In this section, as with the previous work, we will neglect terms of order $v_{wk}^3/\MP$ and only retain first order in $\epsilon$.} with $\imag{\DCz}$, $\imag{\DBarCz}$, $\imag{\phiUz}$, and 
$\imag{\phiDz}$ (the imaginary components of the neutral fields) since 
two linear combinations  of them are eaten by gauge bosons (so there are 
two zero modes).  The four by four mass matrix resulting after the 
spontaneous symmetry breaking can be split into two by two matrices 
for the $\Delta^c$'s and the $\Phi$'s:
\begin{align}
V_{mass}	
	& \supset
	\half 
		\begin{pmatrix}
			\imag{\bar{\Delta}^{c \, 0}}	&
			\imag{\Delta^{c \, 0}}
		\end{pmatrix}
M_{\text{I}\Delta^c}^2
		\begin{pmatrix}
			\imag{\bar{\Delta}^{c \, 0}}	\\
			\imag{\Delta^{c \, 0}}
		\end{pmatrix}
	+ \half
		\begin{pmatrix}
			\imag{\Phi_u^0}			&
			\imag{\Phi_d^0}
		\end{pmatrix}
M_{\text{I}\Phi}^2
		\begin{pmatrix}
			\imag{\Phi_u^0}			\\
			\imag{\Phi_d^0}
		\end{pmatrix}	
\end{align}
where
\begin{equation}
M_{\text{I}\Delta^c}^2 =
\begin{pmatrix}
\frac{Z_A m_{3/2}}{\MP} v^2	& \frac{Z_A m_{3/2}}{\MP} v^2	\\
\frac{Z_A m_{3/2}}{\MP} v^2	& \frac{Z_A m_{3/2}}{\MP} v^2
\end{pmatrix}
\end{equation}
\begin{equation}
M_{\text{I}\Phi}^2 =
\begin{pmatrix}
\half \frac{\kappa_d}{\kappa_u}  \inb{ \frac{Z_\alpha m_{3/2}}{\MP} - \frac{\lambda_A \lambda_\alpha v^2}{\MP^2} } v^2
&
\half \inb{ \frac{Z_\alpha m_{3/2} }{\MP} - \frac{\lambda_A \lambda_\alpha v^2}{\MP^2} } v^2
\\
\half \inb{ \frac{Z_\alpha m_{3/2} }{\MP} - \frac{\lambda_A \lambda_\alpha v^2}{\MP^2} } v^2
&
\half \frac{\kappa_u}{\kappa_d} \inb{ \frac{Z_\alpha m_{3/2}}{\MP} - \frac{\lambda_A \lambda_\alpha v^2}{\MP^2} } v^2
\end{pmatrix}
\end{equation}

The above matrices each have determinant equal to zero, and the remaining 
non-zero eigenvalues are, respectively,
\begin{align}
\label{Eq:Axial Delta Mass}
m_{B^0}^2 &=  \frac{2 Z_A m_{3/2}}{\MP} v^2	\\
\label{Eq:Axial Phi Mass}
m_{A^0}^2 &= \frac{\lambda_\alpha^2 v^4}{2 \MP^2} - 2 m_\Phi^2
\end{align}
where the latter value has been simplified using \eq{sin 2 beta}.  Here 
we have introduced $B^0$ as the axial Higgs boson associated with the 
$\Delta^c$ fields and $A^0$ is the usual MSSM axial Higgs boson.

The mass of $B^0$ will always be positive provided $Z_A > 0$, which 
means that the minus sign in front of the $Z_A$ term in \eq{V non SUSY} 
is crucial for a positive mass-square.  The second could easily be 
positive depending on the value of  $\lambda_\alpha$ and the phase of 
$m_\Phi^2$.

Next we move on to the singly charged fields since they also have two 
zero masses.  The mass matrix for these fields can not be split apart; 
however, if we write
\begin{equation}
V_{mass} \supset
	\begin{pmatrix} 
		\nop{\phiUp}^*	&
		\phiDm		&
		\nop{\DBarCp}^* & 
		\DCm 
	\end{pmatrix}
	M_{SC}^2
	\begin{pmatrix} 
		\phiUp		\\
		\nop{\phiDm}^*	\\
		\DBarCp		\\ 
		\nop{\DCm}^*
	\end{pmatrix}
\end{equation}
then $M_{SC}^2$ can be seen to be three distinct two by two matrices:
\begin{equation}
M_{SC}^2 =
	\begin{pmatrix}
	M_{\Phi^\pm}^2		& \SCmix		\\
	\inp{\SCmix}^\dagger	& M_{\Delta^{c \pm}}^2
	\end{pmatrix}
\end{equation}
where
\begin{equation}
M_{\Phi^\pm}^2 =
\begin{pmatrix}
	  \fourth g_R^2 \kappa_d^2
	+ \half \frac{\kappa_d}{\kappa_u} 
		\inb{
			\frac{Z_\alpha m_{3/2}}{\MP}
			- \frac{\lambda_A \lambda_\alpha v^2}{\MP^2} 
		} v^2
&
	  \fourth g_R^2 \kappa_u \kappa_d
	+ \half \inb{
		\frac{Z_\alpha m_{3/2}}{\MP}
		- \frac{\lambda_A \lambda_\alpha v^2}{\MP^2} 
		} v^2
\\
	  \fourth g_R^2 \kappa_u \kappa_d
	+ \half \inb{ 
		\frac{Z_\alpha m_{3/2}}{\MP}
		- \frac{\lambda_A \lambda_\alpha v^2}{\MP^2} 
		} v^2
&
	\fourth g_R^2 \kappa_u^2
	+ \half \frac{\kappa_u}{\kappa_d} 
		\inb{ 
			\frac{Z_\alpha m_{3/2}}{\MP}
			- \frac{\lambda_A \lambda_\alpha v^2}{\MP^2} 
		} v^2
\end{pmatrix}	
\end{equation}
\begin{equation}
\SCmix =
\begin{pmatrix}
\fourth g_R^2 v \kappa_d 	& - \fourth g_R^2 v \kappa_d		\\
\fourth g_R^2 v \kappa_u	& - \fourth g_R^2 v \kappa_u
\end{pmatrix}
\end{equation}
\begin{equation}
M_{\Delta^{c \pm}}^2 =
\begin{pmatrix}
	\begin{array}{r}
	  \fourth g_R^2 v^2 \inp{1 + \epsilon}
	+ \fourth g_R^2 \inp{\kappa_u^2 - \kappa_d^2}
	\\
	+ \half \inb{ \frac{Z_A m_{3/2}}{\MP} - \frac{\lambda_A^2 v^2}{\MP^2} } v^2
	\end{array}
&
	\begin{array}{r}
	- \fourth g_R^2 v^2
	- \half \inb{ \frac{Z_A m_{3/2} }{\MP} - \frac{\lambda_A^2 v^2}{\MP^2} } v^2
	\\
	\phantom{+ \half \inb{ \frac{Z_A  m_{3/2}}{\MP} - \frac{\lambda_A^2 v^2}{\MP^2} } v^2}
	\end{array}
\\
	- \fourth g_R^2 v^2
	- \half \inb{ \frac{Z_A m_{3/2} }{\MP} - \frac{\lambda_A^2 v^2}{\MP^2} } v^2
&
	\begin{array}{r}
	  \fourth g_R^2 v^2 (1 - \epsilon)
	- \fourth g_R^2 \inp{\kappa_u^2 - \kappa_d^2}
	\\
	+ \half \inb{ \frac{Z_A m_{3/2}}{\MP} - \frac{\lambda_A^2 v^2}{\MP^2} } v^2
	\end{array}
\end{pmatrix}
\end{equation}
Checking the order of magnitude of each of those matrices, it can be 
seen that
\begin{align}
\abs{ \inp{M_{\Phi^\pm}^2}_{ij} } & \sim \epsilon v^2	&
\abs{ \inp{\SCmix}_{ij} } & \sim \sqrt{\epsilon} v^2	&
\abs{ \inp{M_{\Delta^{c \pm}}^2}_{ij} } & \sim v^2
\end{align}
so, $M_{SC}^2$ may be written as
\begin{equation}
\begin{pmatrix}
\epsilon \Lambda_1 v^2			& \sqrt{\epsilon} \Lambda_2 v^2		\\
\sqrt{\epsilon} \Lambda_2^\dagger v^2	& \Lambda_3 v^2
\end{pmatrix}
\end{equation}
where each element of each $\Lambda$ matrix is of order one.  This 
matrix structure is exactly that of the neutrino mass matrix in the Type 
II Singular Seesaw scenario with the associations\footnote{for a review 
of the Type II Singular Seesaw Mechanism see Appendix~\ref{App:Type II Seesaw}}
\begin{align}
\delta^2 m_L 	& \rightarrow M_{\Phi^\pm}^2		&
\delta m_D 	& \rightarrow \SCmix			&
M_R 		& \rightarrow M_{\Delta^{c \pm}}^2
\end{align}

Since the determinant of $M_{\Delta^{c \pm}}^2$ is zero, there is only 
one large eigenvalue given by 
\begin{equation}
m_{D^+}^2 = \half g_R^2 v^2
\end{equation}
the resulting mass matrix for the lighter fields is then read directly 
from the Seesaw formula:
\begin{equation}
\begin{pmatrix}
\fourth g_L^2 \kappa_d^2 
	- \half \frac{\kappa_d}{\kappa_u} 
		\inb{ \frac{Z_\alpha m_{3/2}}{\MP} - \frac{\lambda_A \lambda_\alpha v^2}{\MP^2} } v^2
&
\fourth g_L^2 \kappa_u \kappa_d 
	- \half \inb{ \frac{Z_\alpha m_{3/2}}{\MP} - \frac{\lambda_A \lambda_\alpha v^2}{\MP^2} } v^2
&
0
\\
\fourth g_L^2 \kappa_u \kappa_d 
	- \half \inb{ \frac{Z_\alpha m_{3/2}}{\MP} - \frac{\lambda_A \lambda_\alpha v^2}{\MP^2} } v^2
&
\fourth g_L^2 \kappa_u^2 
	- \half \frac{\kappa_u}{\kappa_d} 
		\inb{ \frac{Z_\alpha m_{3/2}}{\MP} - \frac{\lambda_A \lambda_\alpha v^2}{\MP^2} } v^2
&
0
\\
0
&
0
&
0
\end{pmatrix}
\end{equation}

Evidently zero is one of the eigenvalues, and the determinant of the remaining 
two by two is also zero.  These correspond to the two modes that are eaten by 
the charged gauge bosons.  The trace of the two by two is then the non-zero 
eigenvalue, which corresponds to the MSSM charged Higgs Boson $h^+$. 
So, after using \eq{sin 2 beta}
\begin{equation}
\label{Eq:mhplus}
m_{h^+}^2 = 
	\fourth g_L^2 \kappa^2
	+ \frac{\lambda_\alpha^2 v^4}{2 \MP^2}
	- 2 m_\Phi^2
\end{equation}

 Note that the first term of the right-hand side is just $m_W^2$ and that 
the last two terms sum to the aforementioned $m_{A^0}^2$.  We can 
therefore rewrite \eq{mhplus} as
\begin{equation}
\label{Eq:mhplus2}
m_{h^+}^2 = 
	m_W^2 + m_{A^0}^2
\end{equation}
which matches the MSSM result and will be positive if $m_{A^0}^2$ is.

The remaining charged fields---the doubly charged Higgs bosons---can 
only consist of $\Delta^c$ and $\bar{\Delta}^c$, so this mass matrix is 
a two by two. For these fields we have
\begin{equation}
V_{mass} \supset 
\bigl(
	\begin{array}{cc}
	\nop{\bar{\Delta}^{c\spp}}^*	& \Delta^{c\smm}
	\end{array}
\bigr)
M_{\Delta^{c \pm \pm}}^2 
\Biggl(
	\begin{array}{c}
	\bar{\Delta}^{c\spp}	\\
	\nop{\Delta^{c\smm}}^*
	\end{array}
\Biggr)
\end{equation}
where
\begin{multline}
M^2_{\Delta^{c \pm \pm}} = 
\\
\begin{pmatrix}
	\begin{array}{l}
	  g_R^2 v^2 \epsilon 
	- \half g_R^2 \inp{\kappa_u^2 - \kappa_d^2}
	\\ \; {}
	+ \half \inb{ \frac{Z_A m_{3/2}}{\MP} - \frac{\lambda_A^2 v^2}{\MP^2} }v^2
	+ \frac{\lambda_B \inp{\lambda_A + \lambda_B} v^4 }{\MP^2}
	\end{array}
&
	\inb{\!\inp{
		  \frac{Z_B m_{3/2}}{\MP}
		+ \frac{ \lambda_A \lambda_B v^2 }{\MP^2}
		}
	\! - \! \half \! \inp{
		  \frac{Z_A m_{3/2}}{\MP}
		\! - \! \frac{ \lambda_A^2 v^2}{\MP^2}
	} \! } \! v^2
\\
	\inb{\!\inp{
		  \frac{Z_B m_{3/2}}{\MP}
		+ \frac{ \lambda_A \lambda_B v^2 }{\MP^2}
		}
	\! - \! \half \! \inp{
		  \frac{Z_A m_{3/2}}{\MP}
		\! - \! \frac{ \lambda_A^2 v^2}{\MP^2}
	} \! } \! v^2
&
	\begin{array}{l}
	- g_R^2 v^2 \epsilon 
	+ \half g_R^2 \inp{ \kappa_u^2 - \kappa_d^2 }
	\\ \; {}
 	+ \half \inb{ \frac{Z_A m_{3/2}}{\MP} - \frac{\lambda_A^2 
v^2}{\MP^2} } v^2
	+ \frac{\lambda_B \inp{\lambda_A + \lambda_B} v^4 }{\MP^2}
	\end{array}
\end{pmatrix}
\end{multline}

The eigenvalues can be found, and they are given by
\begin{multline}
m_{D^{\spp}/d^{\spp}}^2 = \\
	\inb{ 
 		    \half \Lambda_A 
		  + \frac{\lambda_B \inp{\lambda_A + \lambda_B} v^2}{\MP^2} 
	}v^2
\pm \sqrt{
 	  \inp{ - g_R^2 \epsilon v^2 + \half g_R^2 \inp{\kappa_u^2 - \kappa_d^2} }^2
	+ \inb{
 		  \Lambda_B 
 		- \half \Lambda_A 
		}^2 v^2
	}
\end{multline}
where
\begin{align*}
\Lambda_A &  \equiv \frac{Z_A m_{3/2}}{\MP} - \frac{\lambda_A^2 v^2}{\MP^2}		&
\Lambda_B &  \equiv \frac{Z_B m_{3/2}}{\MP} + \frac{ \lambda_A \lambda_B v^2 }{\MP^2}
\end{align*}

These eigenvalues are of order $v_{wk}^2$ as expected and are positive for sufficiently large $\lambda_B$.
It is worth noting that a large $\lambda_B$ is consistent with a stable charge conserving vacuum as mentioned earlier.

Finally, we come to the real neutral fields.  These fields, like the singly charged, require use of 
the seesaw mechanism (as discussed in Appendix \ref{App:Type II Seesaw}).  If we write
\begin{equation}
V_{mass} \supset
\half
\bigl(
	\begin{array}{cccc}
	\real{\Phi_u^0}			&
	\real{\Phi_d^0}			&
	\real{\bar{\Delta}^{c \, 0}}	&
	\real{\Delta^{c \, 0}}
	\end{array}
\bigr)
M_{RN}^2
	\begin{pmatrix}
	\real{\Phi_u^0}			\\
	\real{\Phi_d^0}			\\
	\real{\bar{\Delta}^{c \, 0}}	\\
	\real{\Delta^{c \, 0}}		
	\end{pmatrix}
\end{equation}
where
\begin{equation}
M_{RN}^2 = 
\begin{pmatrix}
M_{\Phi \Phi}^2				& M_{\Phi \Delta^c}^2			\\
\inp{M_{\Phi \Delta^c}^2}^\dagger	& M_{\Delta^c \! \bar{\Delta}^c}^2
\end{pmatrix}
\end{equation}
with
\begin{multline}
M_{\Delta^c \! \bar{\Delta}^c}^2 = \\
\begin{pmatrix}
  \half \inp{g_{BL}^2 + g_R^2} v^2 \inp{1 - \epsilon} 
+ \frac{ \lambda_A^2 v^4 }{\MP^2}
&
- \half \inp{g_{BL}^2 + g_R^2} v^2
- \inb{	\frac{Z_A m_{3/2}}{\MP} - \frac{2 \lambda_A^2 v^2}{\MP^2} } v^2
\\
- \half \inp{g_{BL}^2 + g_R^2} v^2
- \inb{	\frac{Z_A m_{3/2}}{\MP} - \frac{2 \lambda_A^2 v^2}{\MP^2} } v^2
&
\half \inp{g_{BL}^2 + g_R^2} v^2 \inp{1 + \epsilon} +  \frac{\lambda_A^2 v^4}{\MP^2}
\end{pmatrix}
\end{multline}
\begin{multline}
M_{\Phi \Phi}^2 =
\\
\begin{pmatrix}
	\fourth \inp{g_L^2 + g_R^2} \kappa_u^2
	+ \half \frac{\kappa_d}{\kappa_u} \!
		\inb{
			\frac{Z_\alpha m_{3/2}}{\MP}
		 	- \frac{\lambda_A \lambda_\alpha v^2}{\MP^2}
		} \! v^2
&
	- \fourth \inp{g_L^2 + g_R^2} \kappa_u \kappa_d
	- \half \! \inb{ \frac{Z_\alpha m_{3/2}}{\MP} -  \frac{\lambda_A \lambda_\alpha v^2}{\MP^2}	} \! v^2
\\
	- \fourth \inp{g_L^2 + g_R^2} \kappa_u \kappa_d
	- \half \! \inb{
		  \frac{Z_\alpha m_{3/2}}{\MP}
		- \frac{\lambda_A \lambda_\alpha v^2}{\MP^2}
		} \! v^2
&
	  \fourth \inp{g_L^2 + g_R^2} \kappa_d^2
	+ \half \frac{\kappa_u}{\kappa_d} \!
		\inb{ \frac{Z_\alpha m_{3/2}}{\MP} -   \frac{\lambda_A \lambda_\alpha v^2}{\MP^2} } \! v^2
\end{pmatrix}
\end{multline}
\begin{equation}
M_{\Phi \Delta^c}^2 =
\begin{pmatrix}
	\frac{1}{2 \sqrt 2} g_R^2 v \kappa_u 
&
	- \frac{1}{2 \sqrt 2} g_R^2 v \kappa_u 
\\
	- \frac{1}{2 \sqrt 2} g_R^2 v \kappa_d 
& 
	\frac{1}{2 \sqrt 2} g_R^2 v \kappa_d 
\end{pmatrix}
\end{equation}
and make the associations
\begin{align}
\delta^2 m_L 	& \rightarrow M_{\Phi \Phi}^2		&
\delta m_D 	& \rightarrow M_{\Phi \Delta^c}^2	&
M_R 		& \rightarrow M_{\Delta^c \! \bar{\Delta}^c}^2
\end{align}
then the single large eigenvalue 
of $M_{\Delta^c \! \bar{\Delta}^c}^2$ is given by 
\begin{equation}
m_{D^0}^2 = \inp{g_{BL}^2 + g_R^2}v^2
\end{equation}

The resulting mass matrix for the lighter fields is
\begin{equation}
\begin{pmatrix}
	\begin{array}{l}
	  \fourth \inb{ g_L^2 + \frac{g_{BL}^2 g_R^2}{g_{BL}^2 + g_R^2} } \kappa_u^2 
	\\ \quad \quad {}
	+ \half \frac{\kappa_d}{\kappa_u} 
		\inb{
			  \frac{Z_\alpha m_{3/2}}{\MP} 
			- \frac{\lambda_A \lambda_\alpha v^2}{\MP^2} 
		}v^2
	\end{array}
&
	\begin{array}{l}
	- \fourth \inb{ g_L^2 + \frac{g_{BL}^2 g_R^2}{ g_{BL}^2 + g_R^2} } \kappa_u \kappa_d 
	\\ \quad \quad {}
	- \half \inb{
			  \frac{Z_\alpha m_{3/2}}{\MP} 
			- \frac{\lambda_A \lambda_\alpha v^2}{\MP^2} 
		}v^2
	\end{array}
&
	0
\\[.75cm]
	\begin{array}{l}
	- \fourth \inb{ g_L^2 + \frac{g_{BL}^2 g_R^2}{g_{BL}^2 + g_R^2} } \kappa_u \kappa_d 
	\\ \quad \quad {}
	- \half \inb{
			  \frac{Z_\alpha m_{3/2}}{\MP} 
			- \frac{\lambda_A \lambda_\alpha v^2}{\MP^2} 
		}v^2
	\end{array}
&
	\begin{array}{l}
	  \fourth \inb{ g_L^2 + \frac{g_{BL}^2 g_R^2}{g_{BL}^2 + g_R^2} } \kappa_d^2 
	\\ \quad \quad {}
	+ \half \frac{\kappa_u}{\kappa_d} 
		\inb{
			  \frac{Z_\alpha m_{3/2}}{\MP} 
			- \frac{\lambda_A \lambda_\alpha v^2}{\MP^2} 
		}v^2
	\end{array}
&
	0
\\[.75cm]
	0
&
	0
&
	\inb{ \frac{3 \lambda_A^2 v^2}{\MP^2} - \frac{Z_A m_{3/2}}{\MP}  }v^2
\end{pmatrix}
\end{equation}

Clearly one of the eigenvalues can be read off, it is
\begin{equation}
m_{d^0}^2 = \inb{ - \frac{Z_A m_{3/2}}{\MP} + \frac{3 \lambda_A^2 v^2}{\MP^2} }v^2
\end{equation}
and is of electroweak order.

The remaining two by two matrix has the eigenvalues
\begin{align}
\label{Eq:real eigenvalues}
m_{H^0/h^0}^2 
	& =	\half
		\inb{
			m_Z^2 + m_A^2 \pm
			\sqrt{
				\inp{m_Z^2 + m_A^2}^2
				-
				4 m_Z^2 m_A^2 \cos{2 \beta}^2
			}
		}
\end{align}
where we have used \eq{Axial Phi Mass} and \eq{m Z} to simplify this 
expression.  Note that these also match MSSM expressions.
%


That completes the Higgs spectrum analysis and we now briefly address additional fermionic content of the theory.  Here we find that there are three light fermions in this model: two doubly charged and a neutral one.  The neutral fermion has a mass in the electroweak range and is the superpartner of the $d^0$.

\subsection{Numerics}

The purpose of this subsection is to validate the above  arguments with 
numerical analysis.  Specifically, our purpose is simply  to show that 
the general arguments about the positivity of the Higgs masses can be 
supported in the parameter space.  Other values of interest are also 
reported including: $v_R$, $\tan{\beta}$, and the difference in the 
ground state values of the CC and CV potentials (as mentioned earlier
we need $\inap{V}_{CV} - \inap{V}_{CC} > 0$, so this is verified in
that last column of \tbl{Vacuum}).

We will keep six of the dimensionful parameters constant (in GeV)
\begin{align*}
m_{\Delta^c} &= 350 		& 
m_{\bar{\Delta}^c} &= 450 	& 
m_{3/2} &= 450			& 
\kappa &= 250 			& 
\MP &= 2.44\E{18}
\end{align*}
and three of the coupling constants at:
\begin{align*}
g_R &= 1.2 & g_L &= .65 & g_1 &= .38
\end{align*}

We vary the remaining as follows:

\begin{center}
\begin{table}[!ht]
\begin{tabular}{|c|c|c|c|c|c|c|c|}
\hline\hline
Case Number	& $\lambda_A$  & $\lambda_B$ & $\lambda_\alpha$ & $Z_A$  & $Z_B$  & $Z_\alpha$  & $m_\Phi^2$(GeV$^2$)	\\
\hline
Case 1		& $0.9$        & $0.8$	    & $0.99$            & $0.65$ & $0.3$  & $1.29$      & $300^2$			\\
Case 2		& $0.5$        & $0.45$	    & $0.1$		& $0.54$ & $0.3$  & $0.16$      & $-100^2$	
\\
Case 3		& $0.4$        & $0.4$	    & $0.2$		& $0.36$ & $0.3$  & $0.29$      & $100^2$	
\\
Case 4		& $0.2$        & $0.3$	    & $0.1$		& $0.18$ & $0.3$  & $0.14$      & $100^2$	
\\
Case 5		& $0.9$        & $0.85$	    & $0.2$		& $0.54$ & $0.3$  & $0.25$      & $-100^2$	
\\
\hline\hline
\end{tabular}
\caption{Points in parameter space used to evaluate the Higg masses}
\label{Table:Parameters}
\end{table}
\end{center}

These values yield the following masses for the Higgs Boson (in GeVs) and the vacuum defining parameters respectively:
\begin{center}
\begin{table}[!ht]
\begin{tabular}{|*{11}{c|}}
\hline\hline
Case Number	& $D^{++}$ & $d^{++}$ & $D^+$        & $H^+$  & $D^0$        & $d^0$ & $H^0$ & $h^0$ & $B^0$ & $A^0$
\\
\hline
Case 1		& $988$     & $161$    & $3.38\E{10}$ & $186$ & $5.01\E{10}$ & $917$ & $167$ & $93$  & $619$ & $167$
\\
Case 2		& $1140$    & $235$    & $4.79\E{10}$ & $188$ & $7.10\E{10}$ & $983$ & $170$ & $90$  & $796$ & $169$
\\
Case 3		& $1170$    & $209$    & $5.23\E{10}$ & $188$ & $7.75\E{10}$ & $953$ & $170$ & $90$  & $718$ & $169$
\\
Case 4		& $1560$    & $555$    & $7.37\E{10}$ & $186$ & $10.9\E{10}$ & $948$ & $167$ & $93$  & $706$ & $167$
\\
Case 5		& $986$     & $166$    & $3.32\E{10}$ & $186$ & $4.91\E{10}$ & $895$ & $167$ & $93$  & $551$ & $167$
\\
\hline\hline
\end{tabular}
\caption{Higgs masses based on parameters from \tbl{Parameters}.  The masses are given in GeV.  As predicted previously, the doubly charged particles ($D^{++}$ and $d^{++}$) have masses in the electroweak range.}
\label{Table:Masses}
\end{table}
\end{center}

\begin{center}
\begin{table}[ht]
\begin{tabular}{|*{5}{c|}}
\hline\hline
Case Number	& $v_R$ (GeV)	& $\epsilon$ 	& $\tan{\beta}$ 	& $\inap{V}_{CV} - \inap{V}_{CC}$ (GeV$^4$)	\\
\hline
Case 1		& $2.8\E{10}$	& $5.0\E{-17}$	& $\infty$		& $1.7\E{27}$					\\
Case 2		& $4.0\E{10}$	& $2.5\E{-17}$	& $9.9$			& $4.0\E{27}$
\\
Case 3		& $4.4\E{10}$	& $2.1\E{-17}$	& $9.9$			& $5.4\E{27}$
\\
Case 4		& $6.1\E{10}$	& $1.0\E{-17}$	& $50$			& $18\E{27}$
\\
Case 5		& $2.8\E{10}$	& $5.2\E{-17}$	& $50$			& $1.6\E{27}$
\\
\hline\hline
\end{tabular}
\caption{Vacuum related parameters based on parameters from \tbl{Parameters}.  The second column shows $v_R$ and as can be seen is the correct order of the seesaw scale.  The last column presents the difference in the ground state energy of the charge violating and the charge conserving vacuum.  A positive value in this column indicates that the charge conserving vacuum is the stable one.}
\label{Table:Vacuum}
\end{table}
\end{center}

\subsection{Implications}

The TeV scale theory in this model differ from MSSM in that we have several new
particles in the $100$ GeV--TeV range.  These particles are : $\Delta^{c++}$,
$\Delta^{c--}$, $d^0$, $\tilde\Delta^{c++}$, $\tilde{\bar\Delta}^{c--}$ and
$\tilde d^0$ where the Higgsinos are differentiated from Higgses by a tildes. 
The charged particles lead to spectacular signatures in colliders due to their
decay modes: $\Delta^{c++} \rightarrow \ell^+ \ell^+$, $\tilde{\Delta}^{c++}
\rightarrow \ell^+ \ell^+ \chi$ ($\chi$ being the lightest neutralino).  On the
other hand, the neutral particles will be hard to produce in the labartory
because of their low coupling values to MSSM matter content.  Their dominant
decay channel is via $d^0 \rightarrow \chi \chi$ with decay lifetimes of the
order $10^{-10}$ sec for generic values of the parameters.  It is worthwile to
mention that $d^0$ and $\tilde d^0$  would have been present in the early
stages of the universe, but would have decayed away before the era of Big Bang 
nucleosynthesis and therefore do not alter our understanding of this period.

\section{Grand unification prospects}
\label{Sec:Grand Unification}

Since the effective TeV scale theory in our model is very different from 
MSSM (due to the presence of a pair of doubly charged fields), it is 
interesting to explore whether there is grand unification of couplings. 
This question was investigated in \cite{dutta1}, where it was noted that if 
there are two pairs of Higgs doublets (corresponding to two bidoublets
$\phi_{1,2}(2,2,0)$), at the TeV scale, the gauge couplings unify around 
$10^{12}$ GeV or so. This raises an interesting point: if there is 
a grand unified theory at $10^{12}$ GeV, then this theory must be 
very different from conventional GUT theories.  The reason this must be so 
is that GUTs violate baryon number and present limits on proton life time 
require that the scale of grand unification be $10^{15}$ GeV. Our GUT theory,
should it exist, must conserve baryon number due to the low unification scale.

An example of such a theory is the $SU(5)\times SU(5)$ model discussed in 
\cite{su5su5}, which embeds the left-right symmetric group we are 
discussing. We do not discuss the details of this theory here, but 
rather indicate the basic features: we envision $SU(5)\times SU(5)$ to be 
broken\cite{su5su5} down to $SU(3)^c\times SU(2)_L\times SU(2)_R\times 
U(1)_{B-L}$ by a Higgs multiplet belonging to the representation 
$\Phi\equiv {\bf (5,\bar{5})}$ with vacuum expectation value (vev) as follows: 
$\inap{\Phi}=\diag{a,a,a,0,0}$.  This is then subsequently broken to the standard 
model.

The fermions in this model belong to the
$({\bf \bar{5},1}) \oplus ({\bf {10},1})\oplus ({\bf 
1, \bar{5}})\oplus ({\bf 1, {10}})$ representation as follows:
\begin{align}
F_L &=
\begin{pmatrix}
D^c \\
D^c \\
D^c \\ 
\nu \\
e
\end{pmatrix}
&
T_L &=
\begin{pmatrix}
0 & U^c & U^c & u & d	\\
-U^c & 0 & U^c & u &d 	\\
-U^c & -U^c & 0 & u & d	\\
-u & -u & -u & 0 & E^+ 	\\
-d & -d & -d & -E^+ & 0
\end{pmatrix} 
\end{align}
and similarly for the right chiral fields.

Implementation of the seesaw mechanism in this model requires that we 
add the Higgs representation 
$({\bf 15, 1})\oplus ({\bf 1,15})$ along with their complex conjugate 
representations. The multiplet b${\bf (1,15)}$ plays the role of 
$\Delta^c$ of the left-right model. When the $\nu^c\nu^c$ component of 
$({\bf 1, 15})$ acquires 
a vev, it gives mass to the right handed neutrino fields triggering the 
seesaw mechanism. The doubly charged Higgs fields are part of the right 
handed $({\bf 1, 15})$ Higgs representation. Symmetry breaking and fermion 
masses in this model are briefly touched on in \cite{su5su5} and will be 
discussed in detail in a separate paper.

\section{Conclusion}

To summarize, we have considered a bottom-up extension of the
MSSM based on the gauge group $SU(2)_L\times SU(2)_R\times U(1)_{B-L}$
that explains small neutrino masses via the seesaw mechanism.  We have also shown 
that if the superpotential of the model is assumed to obey an R-symmetry,
then the $B-L$ breaking scale (seesaw scale) can be predicted to be around 
$10^{11}$ GeV---a phenomenologically acceptable value for this 
scale. This model also solves the $\mu$ problem of the MSSM and predicts two 
TeV scale doubly charged bosons and fermions which 
couple to like sign dileptons and like sign lepton-slepton respectively. 
Such particles have been searched for in various existing experiments 
and will be searched for at the LHC and other future colliders\cite{biswa}. 
Additionally, the model predicts unstable neutral bosons and fermions
which can not be easily probed by experiment, but which would have been 
produced in the early universe.

Finally we note that the conclusions of this paper can be equally applied to
the group $SU(2)_L \times U(1)_{I_{3R}} \times U(1)_{B-L}$, with the mass 
spectrum being identical except for the lack of light doubly charged particles
and a heavy singly charged.

\section{Acknowledgements}

This work is supported by the National Science Foundation
grant no.~Phy-0354401.

\appendix

\section{Type II Singular Seesaw Mechanism}
\label{App:Type II Seesaw}

We start with a mass matrix of the form
\begin{equation}
\mathscr{M} =
\begin{pmatrix}
\delta^2 m_L		& \delta m_D	\\
\delta m_D^\dagger	& M_R
\end{pmatrix}
\end{equation}
where $\delta$ carries the relative order of magnitude of the elements of each of the three $(n \times n)$ matrices $m_L$, $m_D$, and $M_R$---i.e.~there is a hierarchy which can be thought of as either 
$\abs{\inp{M_R}_{ij}} \gg \abs{\inp{\delta m_D}_{k \ell}} \gg \abs{\inp{\delta^2 m_L}_{pq}}$; or, alternatively, $\delta \ll 1$, $\abs{\inp{M_R}_{ij}} \sim \abs{\inp{m_D}_{k\ell}} \sim \abs{\inp{m_L}_{pq}} \equiv v$.

It is not assumed, however, that all the eigenvalues of $M_R$ are of this high scale $v$, so in the limit $\delta \rightarrow 0$, it is possible that $\det M_R = 0$.  Therefore, to exploit this hierarchy it is necessary to extract those smaller eigenvalues.  This is done as follows:

First, diagonalize $M_R$ through an $(n \times n)$ rotation $R$ via
\begin{equation}
\mathscr{R} = 
\begin{pmatrix}
\mathbbm{1}	& 0	\\
0		& R
\end{pmatrix}
\end{equation}
so that
\begin{equation}
\mathscr{R} \mathscr{M} \mathscr{R}^T =
\begin{pmatrix}
\delta^2 m_L		& \delta m_D R^T	\\
\delta R m_D^\dagger	& M_d	
\end{pmatrix}
\end{equation}
with $M_d \equiv R M_R R^T$ which is a diagonal matrix.  The matrix $R$ should be chosen so that the first $k$ eigenvalues of $M_d$ are the small ones---thus, for $1 \le i \le k$
\begin{equation}
\inp{\delta^2 \mu_R}_{ii} \equiv \inp{M_d}_{ii} = \delta^2 \lambda_{i} v^2
\end{equation}
where $\lambda_i \sim 1$ and $\inp{\delta^2 \mu_R}_{ij} = 0$ for $i \ne j$.  The remaining (large) eigenvalues are then placed in a separate matrix:
\begin{equation}
\Delta_R \equiv \diag{ \inp{M_d}_{k+1,k+1}, \inp{M_d}_{k+2,k+2}, \ldots, \inp{M_d}_{n,n} }
\end{equation}

Also define
\begin{equation}
\delta \mu_1 \equiv 
\begin{pmatrix}
0					& \col_1\inp{\delta m_D R^T}	& \cdots	& \col_k\inp{\delta m_D R^T} \\
\row_1\inp{\delta R m_D^\dagger}	& 0				& \cdots	& 0			     \\
\vdots					& \vdots			& \ddots	& \vdots		     \\
\row_k\inp{\delta R m_D^\dagger}	& 0				& \cdots	& 0			     \\
\end{pmatrix}
\end{equation}
\begin{equation}
\delta^2 \mu_2 \equiv
\begin{pmatrix}
\delta^2 m_L	& 0			\\
0		& \delta^2 \mu_R
\end{pmatrix}
\end{equation}
\begin{equation}
\delta \mu_D \equiv
\begin{pmatrix}
\col_{k+1}\inp{\delta m_D R^T}	& \col_{k+2}\inp{\delta m_D R^T}	& \cdots	& \col_{n}\inp{\delta m_D R^T}
\\
0				& 0					& \cdots	& 0
\end{pmatrix}
\end{equation}

With those definitions we may write
\begin{equation}
\mathscr{R} \mathscr{M} \mathscr{R}^T = 
\begin{pmatrix}
\delta \mu_1 + \delta^2 \mu_2	& \delta \mu_D	\\
\delta \mu_D^\dagger		& \Delta_R
\end{pmatrix}
\end{equation}

Now a matrix $P$ is chosen so that it block diagonalizes $\mathscr{R} \mathscr{M} \mathscr{R}^T$.  This $P$ is implemented through $\mathscr{P}$ which, to order $\delta^2$, is given by
\begin{equation}
\mathscr{P} =
\begin{pmatrix}
\mathbbm{1} - \half \delta^2 P P^\dagger	& - \delta P					\\
\delta P^\dagger				& \mathbbm{1} - \half \delta^2 P^\dagger P
\end{pmatrix}
\end{equation}

The matrix $P$ is then determined by the requirement
\begin{equation}
\mathscr{P} \mathscr{R} \mathscr{M} \mathscr{R}^T \mathscr{P}^\dagger =
\begin{pmatrix}
m		& 0		\\
0		& M
\end{pmatrix}
\end{equation}
with $m$ the $(n + k) \times (n + k)$ mass matrix of interest and $M = \Delta_R + \mathcal{O}\inp{\delta}$.  The off-block-diagonal condition yields
\begin{equation}
P = \mu_D \Delta_R^{-1}
\end{equation}
and then using that $P$, the mass matrix for the light eigenstates can be determined:
\begin{equation}
m = \delta \mu_1 + \delta^2 \mu_2 - \delta^2 \mu_D \Delta_R^{-1} \mu_D^\dagger
\end{equation}



\end{document}